\documentclass[aps,prl,reprint,floatfix,superscriptaddress,nofootinbib]{revtex4-2}
\usepackage{amsmath,amssymb,graphicx,array,booktabs,multirow,hyperref}
\hypersetup{hidelinks}
\newcommand{\jpsi}{J/\psi}
\newcommand{\cccc}{cc\bar c\bar c}
\newcommand{\bbbb}{bb\bar b\bar b}
\begin{document}
\title{Finite-Size Heavy-Diquark Clusters in Fully Heavy Tetraquarks and the Radial Structure of X(6600), X(6900) and X(7100)}
\author{Guo-Fa Yang}
\affiliation{School of Science, Shenzhen Campus of Sun Yat-sen University, Shenzhen 518107, China}
\author{Liu-Yu Zhang}
\affiliation{School of Science, Shenzhen Campus of Sun Yat-sen University, Shenzhen 518107, China}
\author{Tian-Wei Wu}
\email{wutw6@mail.sysu.edu.cn}
\affiliation{School of Science, Shenzhen Campus of Sun Yat-sen University, Shenzhen 518107, China}
\begin{abstract}
A heavy diquark is a useful composite constituent only if its internally determined properties remain predictive in different surroundings. We extend a finite-size framework established for doubly heavy baryons to two interacting heavy diquark clusters and investigate the $2^{++}$ spectra of $\cccc$ and $\bbbb$. Two established quark potentials determine the cluster masses and peak-height-weighted scales without fitting tetraquark levels. The spectra and four-body wave-function projections support interpreting $X(6600)$, $X(6900)$, and $X(7100)$ as the intercluster $1S$, $2S$, and $3S$ sequence, with an additional predominantly internal excitation near $6.98$ GeV. The latter lies between successive intercluster excitations, showing why four-body level order does not fix radial assignments. Despite the larger bottom mass, ground-state spatial-resolution measures remain close to those in the charm sector: approximately $25$--$26\%$ versus $27\%$. Relative internal and intercluster scales therefore control the finite-size effect. These results test the transferability of a finite-size heavy-quark cluster and connect its spatial structure to the organization of multiquark spectra.

\end{abstract}
\maketitle
The discovery of hadrons beyond the conventional meson and baryon assignments of the quark model has turned multiquark spectroscopy into an experimental subject. Charged charmoniumlike states, doubly charmed tetraquarks, and hidden-charm pentaquarks~\cite{BESIII:2013ris,LHCb:2014zfx,tcc2022,penta2015,LHCb:2019kea,LHCb:2022ogu} demonstrate the variety of heavy-quark environments that a structural description must address~\cite{Liu:2019zoy,Chen:2022asf,Guo:2017jvc,Johnson:2024omq,Zhang:2025vew,Ma:2017nik,Wu:2022gie}. Fully heavy tetraquarks offer a particularly direct test of the organization of heavy constituents: without light valence quarks, their internal and intercluster length scales both emerge from heavy-quark dynamics. They therefore probe whether a heavy diquark remains a useful composite constituent when its surroundings are also heavy and spatially extended.

The experimental developments sharpen this question. LHCb observed the $X(6900)$ structure in the di-$\jpsi$ spectrum~\cite{LHCb:2020bwg}; ATLAS and CMS subsequently reported further dicharmonium structures~\cite{ATLAS:2023bft,CMS:2023owd}. The CMS analyses of $X(6600)$, $X(6900)$, and $X(7100)$ suggest a common family~\cite{CMS:2026tiu}, while an angular analysis favors $J^{PC}=2^{++}$ under its assumptions of common quantum numbers and couplings~\cite{CMS:2025fpt}. These observations motivate an $S$-wave system of two spin-1 clusters. They leave open a more specific spectroscopic question: are the observed structures successive radial excitations between the clusters, or do excitations within a cluster enter the same energy region? A sequence of fitted masses does not by itself distinguish these possibilities.

Several theoretical approaches already place fully heavy configurations in this region. Constituent four-body calculations~\cite{Liu:2019zuc,liu:2020eha,Lu:2020cns,Zhang:2022qtp,Wang:2019rdo}, Gaussian-expansion studies~\cite{Wang:2022yes,Chen:2020lgj}, diffusion Monte Carlo~\cite{Gordillo:2020sgc}, and relativized spectroscopy~\cite{Yu:2022lak} explore the resolved dynamics, with cross-model comparisons provided in Ref.~\cite{Wu:2024euj}. Diquark--antidiquark descriptions instead organize the spectrum through composite constituents~\cite{Debastiani:2017msn,Bedolla:2019zwg,Faustov:2022mvs,Zhu:2020xni}. Color-magnetic interactions~\cite{Yan:2023lvm}, QCD sum rules~\cite{Wang:2018poa,Chen:2016jxd,Agaev:2023wua}, multiquark static potentials~\cite{Asadi:2021ids}, and lattice-inspired interactions~\cite{Yang:2021hrb} provide complementary perspectives; production calculations also offer spin-sensitive information~\cite{Zhang:2020hoh}. The variety of nearby configurations makes excitation content essential. A useful connection between the four-body and cluster descriptions must explain which motion produces a level, as well as where that level lies.

Quark clustering reduces the complexity of multiquark dynamics~\cite{Anselmino:1992vg,Maiani:2004vq}, but its utility depends on whether a cluster carries consistent properties across states and environments. Heavy antiquark--diquark symmetry (HADS) supplies the point-source limit for a compact color-antitriplet heavy pair~\cite{savage1990,brambilla2005,fleming2006,hu2006,mehen2006}. At physical heavy-quark masses, retaining the diquark mass does not retain its spatial extent. A finite-mass point-source treatment can consequently overbind doubly heavy baryons: the surroundings resolve an extended pair rather than a coincident color source. Our previous study~\cite{Yang:2026uif} addressed this missing contribution by assigning the heavy diquark both a mass and a representative size determined from its isolated dynamics. The same scale prescription describes ground and radial excitations in nonstrange and strange doubly heavy baryons and quantifies their state-dependent finite-size HADS corrections. Extending that prescription to two finite heavy sources tests the transferability of the constituent itself.

Two routes to an effective cluster interaction can thus be distinguished. A point-particle description can absorb unresolved structure phenomenologically into readjusted effective couplings. For example, the point-diquark spectrum of Ref.~\cite{Zhu:2020xni}, used in the CMS comparison~\cite{CMS:2026tiu}, adopts $m_D=3.0$ GeV and $\alpha_s=0.35$ and assigns masses $6566$, $6890$, and $7160$ MeV to the $2^{++}$ levels with $n=2,3,4$. Its coupling differs from the underlying charmonium-based input~\cite{Debastiani:2017msn}; the radial correspondence alone therefore does not isolate a finite-size effect. The alternative developed in Ref.~\cite{Yang:2026uif} retains the source size explicitly and inherits the interaction parameters from the quark-pair description. This makes a three-body-to-four-body extension possible without fitting a new coupling to each hadronic environment. Here we test that extension, compare the point and finite-size limits at common cluster inputs, and use four-body wave-function projections to identify the excitation modes.

We study the $S$-wave $J^{PC}=2^{++}$ sector of $QQ\bar Q\bar Q$ ($Q=c,b$) in the color-singlet $\bar{\mathbf3}_c\otimes\mathbf3_c$ channel. Pauli symmetry fixes each internal cluster spin to one; the other allowed couplings, $J=0,1$, are outside the present scope. Model I is the OGE-plus-linear interaction of Ref.~\cite{Liu:2019zuc}, and Model II is AL1~\cite{Semay:1994ht,Silvestre-Brac:1996myf}, both with their published quark parameters and no tetraquark refit.

Gaussian expansion~\cite{Kamimura:1988zz,Hiyama:2003cu} first determines the isolated axial-vector diquark eigenstates $\phi_a$, masses $m_{D,a}=2m_Q+\epsilon_a$, and normalized radial probabilities $P_a(r)$. For their local maxima at $r_i$, the representative separation is

\begin{equation}
 r_{QQ,a}=\frac{\sum_i P_a(r_i)r_i}{\sum_iP_a(r_i)}.
 \label{eq:scale}
\end{equation}
For the ground state this is the most-probable separation; for an excitation it includes all radial peaks. The prescription is inherited from the baryon study~\cite{Yang:2026uif}. Cluster masses and sizes are fixed by the isolated spin-1 pair before solving the tetraquark problem and remain unchanged along an intercluster excitation sequence.

We label a finite-size state by $(a,b,n)=(n_{\rho_1},n_{\rho_2},n_\lambda)$, where each index counts radial nodes starting at zero. Thus $(0,0,0)$ is the ground state, $(0,0,1)$ and $(0,0,2)$ are the first two intercluster excitations, and $(1,0,0)$ excites one cluster internally. For $C=+$ the latter denotes the normalized symmetric combination with $(0,1,0)$; these exchanged labels do not represent two distinct levels. The coordinates are $\mathbf r_{\rho_1}=\mathbf r_1-\mathbf r_2$, $\mathbf r_{\rho_2}=\mathbf r_3-\mathbf r_4$, and $\mathbf r_\lambda=(\mathbf r_1+\mathbf r_2-\mathbf r_3-\mathbf r_4)/2$. 

\begin{figure}[t]
\centering\includegraphics[width=.6\columnwidth]{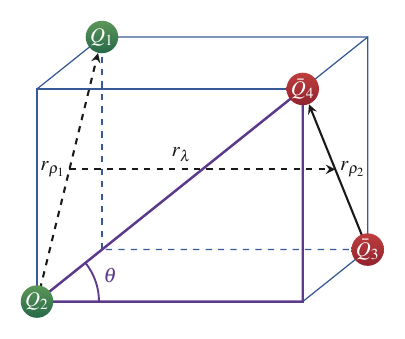}
\caption{Internal displacements $\mathbf r_{\rho_1}$, $\mathbf r_{\rho_2}$ and intercluster displacement $\mathbf r_\lambda$ in the representative geometry. The construction defines the finite-size interaction and the spatial-resolution measures.}
\label{fig:geometry}
\end{figure}

In the representative geometry of Fig.~\ref{fig:geometry}, the three vectors are mutually perpendicular. The four cross-pair distances then reduce to a common separation,

\begin{equation}
 s_{ab}(r_\lambda)=\sqrt{r_\lambda^2+\frac{r_{QQ,a}^2+r_{\bar Q\bar Q,b}^2}{4}}.
 \label{eq:geometry}
\end{equation}
The finite-size central interaction is obtained by evaluating the color-singlet cluster potential at $s_{ab}$; the geometry itself assumes no particular potential. Both models use composite cluster masses for recoil and contact strength, while the Gaussian spin kernel retains its quark-mass-dependent range and is normalized in $r_\lambda$~\cite{Wang:2022clw,Beinker:1995qe}. This specifies an effective-cluster prescription, not an exact folding of the six pair interactions.

The corresponding point approximation sets both radii to zero while keeping the diquark masses $M_1,M_2$, the recoil, and the contact prescription unchanged. Thus the comparison isolates the central finite-size interaction within this cluster model. Since the Cornell-type central terms increase with distance, $s_{ab}\geq r_\lambda$ raises the radial eigenvalues relative to the point limit. In the point model an internal excitation primarily changes the cluster rest masses and recoil; the finite-size model also retains the enlarged source. This distinction predicts a particularly large correction for an internally excited state.

\begin{figure}[t]
\centering\includegraphics[width=.7\columnwidth]{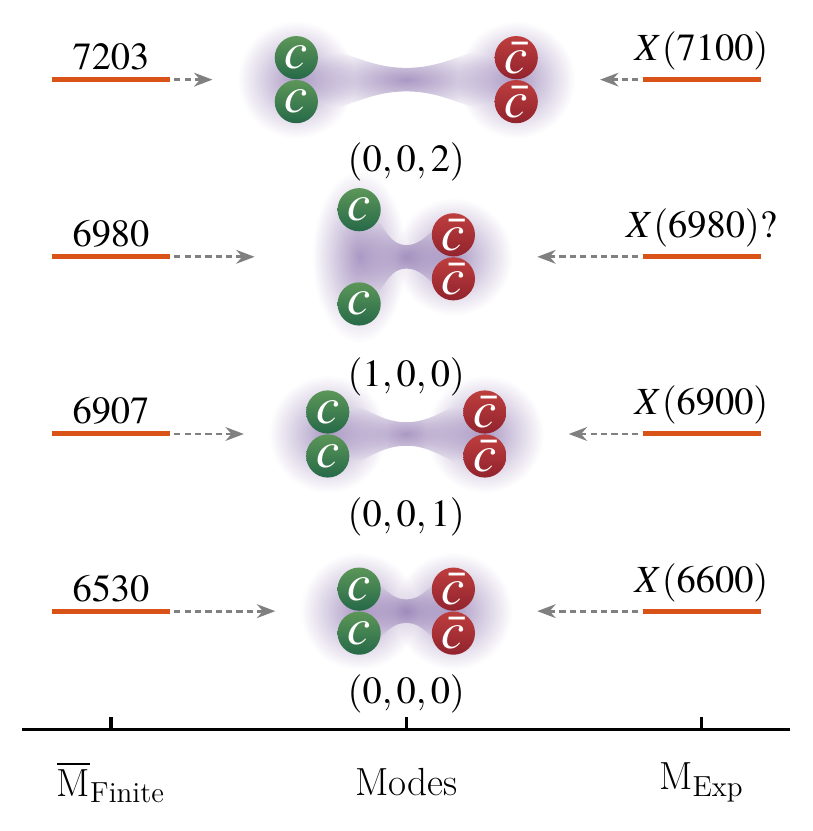}
\caption{Finite-size fully charmed spectrum and proposed excitation assignments. Masses are averages of Models I and II. $X(6980)?$ denotes the predicted internal-excitation candidate near $6980$ MeV; the other $X$ labels name observed structures rather than their exact fitted masses.}
\label{fig:assignment}
\end{figure}

Table~\ref{tab:structure} collects the finite-size masses and structural observables. For $\cccc$, the $(0,0,0)$, $(0,0,1)$, and $(0,0,2)$ masses are $6542$, $6924$, and $7218$ MeV in Model I and $6518$, $6889$, and $7188$ MeV in Model II. Their placement supports the intercluster $1S,2S,3S$ interpretation of $X(6600)$, $X(6900)$, and $X(7100)$ (Fig.~\ref{fig:assignment}). The model means, approximately $6530$, $6907$, and $7203$ MeV, differ from the CMS coherent-fit central values~\cite{CMS:2026tiu} by $63$, $60$, and $30$ MeV. The internal excitation $(1,0,0)$ lies at $6952$--$7007$ MeV, with mean $6980$ MeV, between the first two intercluster excitations. Resolving internal dynamics therefore adds a nearby candidate rather than simply relabeling the radial sequence.

\begin{table}[t]
\caption{Masses $M_{\rm FS}$, structure-resolution angles $\theta$, state-dependent $\delta_{\mathrm{HADS}}$, and intercluster rms radii of the $J^{PC}=2^{++}$ in Models~I and II. The super/subscript FS denotes the finite-size calculations. Masses are in MeV, $\theta$ in degrees, $\delta_{\mathrm{HADS}}$ in percent, and radii in fm.}
\centering
\renewcommand{\arraystretch}{0.96}
\begin{tabular}{c c c c c c}
\hline\hline
Model
& $(n_{\rho_1},n_{\rho_2},n_\lambda)$
& $M_{\rm FS}$
& $\theta$
& $\delta_{\text{HADS}}$
& $r_{\lambda,\text{rms}}^{\rm FS}$ \\
\hline
\multicolumn{6}{c}{$cc\bar{c}\bar{c}$} \\
\hline
 I  & (0, 0, 0) & 6542 & 42 & 26.9 & 0.41 \\
    & (0, 0, 1) & 6924 & 29 & 15.6 & 0.73 \\
    & (1, 0, 0) & 7007 & 51 & 37.9 & 0.45 \\
    & (0, 0, 2) & 7218 & 23 & 11.4 & 1.00 \\
\hline
 II & (0, 0, 0) & 6518 & 41 & 26.5 & 0.40 \\
    & (0, 0, 1) & 6889 & 29 & 16.0 & 0.69 \\
    & (1, 0, 0) & 6952 & 50 & 37.0 & 0.43 \\
    & (0, 0, 2) & 7188 & 24 & 11.8 & 0.94 \\
\hline
\multicolumn{6}{c}{$bb\bar{b}\bar{b}$} \\
\hline
 I  & (0, 0, 0) & 19374 & 40 & 25.2 & 0.23 \\
    & (0, 0, 1) & 19722 & 25 & 12.6 & 0.45 \\
    & (1, 0, 0) & 19813 & 50 & 37.0 & 0.28 \\
    & (0, 0, 2) & 19957 & 20 & 8.7  & 0.64 \\
\hline
 II & (0, 0, 0) & 19360 & 40 & 25.8 & 0.23 \\
    & (0, 0, 1) & 19705 & 26 & 13.2 & 0.43 \\
    & (1, 0, 0) & 19790 & 50 & 37.2 & 0.27 \\
    & (0, 0, 2) & 19950 & 20 & 9.2  & 0.61 \\
\hline\hline
\end{tabular}
\label{tab:structure}
\end{table}

The assignment is supported by a wider scan in which all three radial indices run independently from zero to three. After identifying exchanged internal labels, each model contains 40 distinct $C$-even reference configurations. Both primary interactions place $(0,0,0)$, $(0,0,1)$, $(1,0,0)$, and $(0,0,2)$ first in the charm and bottom spectra. In charm, the next mixed excitation $(1,0,1)$ is at $7279$--$7333$ MeV, while $(2,0,0)$ is at $7295$--$7361$ MeV. The low-lying pattern is therefore not selected by omitting a cheaper internal mode. The first internal excitation costs about $434$--$465$ MeV, compared with $371$--$383$ MeV for the first intercluster excitation; subsequent steps decrease along each unmixed sequence. Internal excitation changes the source mass, size, and recoil together, so its cost cannot be inferred from the isolated diquark spacing alone.

The fixed-input comparison between the point approximation and finite-size shows that the later changes both masses and excitation order. In charm the ground state rises by $307$--$421$ MeV and the internally excited state by $450$--$595$ MeV, whereas the first two intercluster excitations rise by only $172$--$207$ and $130$--$151$ MeV. Bottom follows the same hierarchy. The point approximation places $(1,0,0)$ below $(0,0,1)$, while both finite-size and four-body calculations reverse that order. An enlarged internal source thus contributes physics that cannot be supplied by its excitation energy alone, or isolated by comparing calculations with different effective couplings.

The four-body comparison retains all six pair interactions and three coupled radial coordinates in one $H$-type Jacobi configuration, with $l_{\rho_1}=l_{\rho_2}=l_\lambda=0$. Pauli symmetry selects the color $\bar{\mathbf3}_c\otimes\mathbf3_c$ channel in this spin and spatial sector~\cite{Liu:2019zuc,Wang:2019rdo}. The lowest levels $k=1,\ldots,4$, labeled $1S$--$4S$ in the figures, specify energy order rather than pure intercluster excitations.

To resolve their excitation content, we project all four eigenstates onto a common orthonormal product basis. The internal functions $\phi_a$ are isolated-pair eigenstates. The external functions $\widetilde\chi_n$ use the ground-state finite-size interaction throughout, with the four-quark Jacobi reduced mass $m_Q$ rather than the composite mass used for the cluster spectrum. Suppressing the common color--spin factor, the normalized $C$-even reference state and its weight are

\begin{align}
 \Phi^+_{abn}&=\frac{\phi_a(r_{\rho_1})\phi_b(r_{\rho_2})+\phi_b(r_{\rho_1})\phi_a(r_{\rho_2})}{\sqrt{2(1+\delta_{ab})}}
 \widetilde\chi_n(r_\lambda),\nonumber\\
 p_{abn}^{(k)}&=|\langle\Phi^+_{abn}|\Psi_k\rangle|^2.
 \label{eq:projection}
\end{align}
We use $a\geq b$ to avoid double counting. $P_{\rm main}$ is the dominant weight, $P_{\rm minor}$ the specified subdominant weight, and $P_{\rm tail}=1-P_{\rm main}-P_{\rm minor}$ includes all other components. The overlaps are evaluated coherently in the Gaussian basis.

\begin{figure}[t]
\centering\includegraphics[width=\columnwidth]{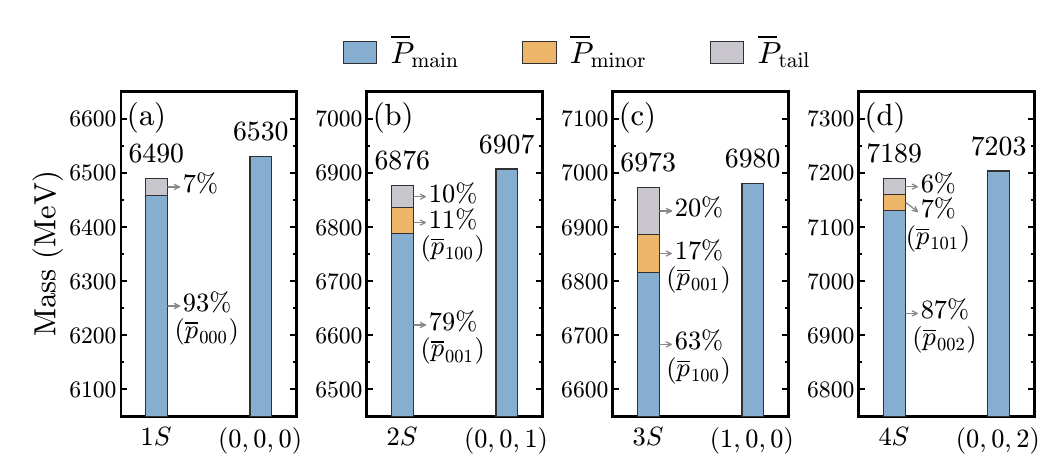}
\caption{Four-body and finite-size masses and radial content in $\cccc$, averaged over Models I and II. The four-body labels $1S$--$4S$ denote energy order; the triples denote the corresponding dominant reference modes. }
\label{fig:charm_projection}
\end{figure}

Figure~\ref{fig:charm_projection} identifies the sequence $(0,0,0)\to(0,0,1)\to(1,0,0)\to(0,0,2)$. For charm, the dominant weights are $92$--$93\%$, $78$--$81\%$, $61$--$65\%$, and $85$--$88\%$. The third level also contains $15$--$18\%$ of $(0,0,1)$, while the second contains $10$--$12\%$ of $(1,0,0)$. The neighboring excited levels therefore mix internal and intercluster motion, even though their largest components remain distinguishable. The proposed observed sequence corresponds to the first, second, and fourth four-body levels. 

The spatial moments supply an independent check of this interpretation. In Model II, the second and third charm levels have intercluster rms radii $0.63$ and $0.47$ fm, respectively, but internal rms separations $0.53$ and $0.66$ fm. Model I gives the same reversal: the intercluster radius decreases from $0.67$ to $0.52$ fm while the internal separation increases from $0.57$ to $0.71$ fm. The higher of these levels is thus more internally extended but externally more compact, as expected for a change of dominant mode. These moments are computed from the four-body wave functions and are not the isolated peak-height-weighted inputs.

\begin{figure}[t]
\centering\includegraphics[width=\columnwidth]{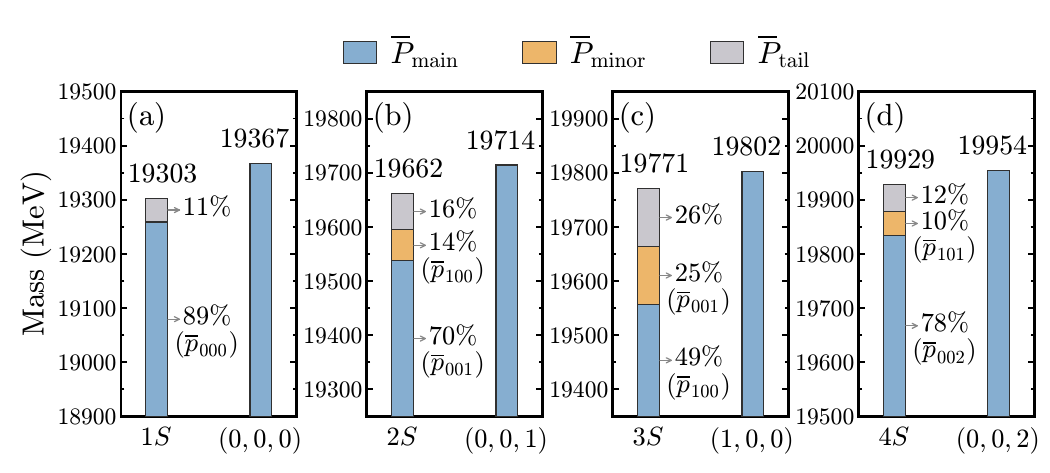}
\caption{The corresponding four-body/finite-size mass comparison and radial decomposition in $\bbbb$. The labeling and conventions are those of Fig.~\ref{fig:charm_projection}. The internally excited third level has stronger mixing than its charm counterpart.}
\label{fig:bottom_projection}
\end{figure}

In bottom the same sequence persists (Fig.~\ref{fig:bottom_projection}). The finite-size intercluster masses average $19367$, $19714$, and $19954$ MeV, with an intervening internal excitation near $19802$ MeV. The third level has only $48$--$50\%$ internal-excitation weight, accompanied by $25$--$26\%$ of $(0,0,1)$; a larger heavy mass therefore does not eliminate mode mixing. Its four-body internal rms separation, $0.386$--$0.398$ fm, exceeds that of the second level, $0.325$--$0.337$ fm, while its intercluster radius is smaller. The same structural comparison works in both heavy-flavor sectors.

The mass residuals quantify the improvement without being used to fix the cluster inputs. Relative to the mode-matched four-body levels, every point-source mass lies low, by $97$--$764$ MeV across the 16 comparisons. The finite-size residuals instead range from $-4$ to $+66$ MeV; their mean absolute value is about $33$ MeV, compared with $326$ MeV for the point approximation. For charm the four-body masses average $6490$, $6876$, $6973$, and $7189$ MeV, close to the finite-size means in Fig.~\ref{fig:charm_projection}. This agreement tests the utility of the inherited scale across different modes.

The same framework quantifies the state dependence of spatial resolution. Extending the structure angle of Ref.~\cite{Yang:2026uif} to two extended sources gives

\begin{align}
 \hat\theta_{ab}(r_\lambda)&=\arctan\frac{\sqrt{r_{QQ,a}^2+r_{\bar Q\bar Q,b}^2}}{2r_\lambda},\nonumber\\
 \theta&=\langle\hat\theta_{ab}\rangle,\qquad
 \delta_{\rm HADS}=1-\langle\cos\hat\theta_{ab}\rangle.
 \label{eq:delta}
\end{align}
The expectation values are evaluated in the finite-size intercluster state, not in the projection reference. The operator vanishes in the point-source limit and grows when the source becomes large relative to the sampled intercluster distance. Its definition is geometric and contains no explicit heavy-mass factor; the mass dependence enters through the internal scales and external dynamics. We retain the notation of the baryon study to quantify the finite-size departure from the point-source description. It does not represent the total breaking of all HADS relations, nor a universal conversion of a geometric correction into an energy shift.

\begin{figure}[t]
\centering\includegraphics[width=.96\columnwidth]{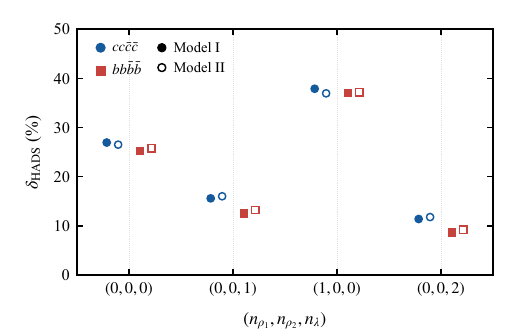}
\caption{State-dependent spatial-resolution measure. Blue circles and red squares denote $\cccc$ and $\bbbb$; filled and open symbols denote Models I and II. Internal and intercluster excitations produce opposite trends.}
\label{fig:delta}
\end{figure}

The contrasting trends in Fig.~\ref{fig:delta} follow from this relative geometry. With both clusters internally unexcited, $\delta_{\rm HADS}$ in charm decreases from $26.5$--$26.9\%$ for $(0,0,0)$ to $15.6$--$16.0\%$ for $(0,0,1)$ and $11.4$--$11.8\%$ for $(0,0,2)$. The unchanged sources are sampled at larger distances. Exciting one source instead raises the measure to $37.0$--$37.9\%$. Bottom exhibits the same opposition, with $25.2$--$25.8\%$, $12.6$--$13.2\%$, and $8.7$--$9.2\%$ along the intercluster sequence and $37.0$--$37.2\%$ for $(1,0,0)$. The scan continues these trends along the unmixed paths: for $(0,0,3)$ the values fall to $9.2$--$9.5\%$ in charm and $6.8$--$7.3\%$ in bottom, whereas higher internal excitations increase the resolution. A small source size in absolute units is therefore insufficient to justify the point approximation.

The ground-state charm--bottom comparison is particularly instructive. The isolated separation decreases from $0.40$--$0.42$ fm for $cc$ to $0.22$ fm for $bb$, but the finite-size intercluster rms radius simultaneously decreases from $0.40$--$0.41$ fm to $0.23$ fm. Consequently the geometric resolution remains near $27\%$ and $25$--$26\%$, respectively. A heavier source becomes smaller, but the second heavy cluster probes it from a smaller distance. Comparing these systems approximately controls the relative geometry while changing the heavy mass, helping to distinguish geometric resolution from heavy-flavor dependence. It does not remove the dynamical connection between mass and size or establish the asymptotic heavy-mass limit. In a light-spectator environment the external length instead remains governed mainly by light dynamics, explaining why the baryon and fully heavy systems need not approach the point-source regime at the same rate.

The resulting interpretation is a spectroscopic one. The four-body projections resolve mode mixing within the stated radial space; scattering channels and resonance poles require further dynamics. In particular, the nearby internal candidate may mix with the intercluster excitation associated with $X(6900)$ and need not appear as an isolated peak. Measurements in di-$\jpsi$ and $\jpsi\psi(2S)$~\cite{CMS:2026tiu,ATLAS:2026ratio}, together with spin-sensitive observables, provide tests beyond mass matching. A joint description of their line shapes and branching fractions will require transition amplitudes and open-channel effects. The distinct internal and intercluster modes identified here supply a physical basis for that next step.

The extension from doubly heavy baryons to fully heavy tetraquarks supports a heavy-diquark constituent whose mass and size are supplied by its own dynamics and whose interaction can be transferred without refitting couplings. The same finite-size prescription accommodates the intercluster $1S,2S,3S$ assignment of $X(6600)$, $X(6900)$, and $X(7100)$ and an intervening internal excitation near $6980$ MeV. Four-body projections and radii explain why the intercluster sequence skips a level in the resolved spectrum and identify the mixing that limits a single-mode description. The comparable spatial resolution in charm and bottom further shows that relative cluster geometry, as well as finite heavy mass, controls the departure from a point source. These linked spectral and structural tests give concrete physical meaning to a finite-size heavy-quark cluster as an effective constituent of more complex hadrons.

\begin{acknowledgments}
This work is supported by the National Natural Science Foundation of China under Grant No.~12405108. Tian-Wei Wu acknowledges support from the Appointed Position of Pengcheng Peacock Plan (C Track), Shenzhen.

\end{acknowledgments}
\bibliographystyle{apsrev4-2}
\bibliography{references}
\end{document}